# On the Stochastic Origin of Quantum Mechanics

Roumen Tsekov

Department of Physical Chemistry, University of Sofia, 1164 Sofia, Bulgaria

The quantum Liouville equation, which describes the phase space dynamics of a quantum system of fermions, is analyzed from stochastic point of view as a particular example of the Kramers-Moyal expansion. Quantum mechanics is extended to relativistic domain by generalizing the Wigner-Moyal equation. Thus, an expression is derived for the relativistic mass in the Wigner quantum phase space presentation. The diffusion with an imaginary diffusion coefficient is discussed. An imaginary stochastic process is proposed as the origin of quantum mechanics.

One of many existing interpretations of quantum mechanics is the stochastic one, which is summarized in the seminal Nelson paper [1]. The Schrödinger equation is derived there from a real Wiener process, but the Nelson approach says nothing about the evolution of the quantum probability in the momentum space, which is compulsory for the complete mechanical treatment. Moreover, it is well known that the instant velocity of a Wiener process is infinite and thus the Nelson description is not equivalent to quantum mechanics. Obviously, the correct stochastic analysis requires consideration in the phase space. In the phase space formulation of quantum mechanics, the Schrödinger equation transforms to the quantum Liouville (Wigner-Moyal) equation [2-4]

$$\partial_t W + p \cdot \partial_q W / m = \sum_{n=0}^{\infty} \frac{(i\hbar/2)^{2n}}{(2n+1)!} \partial_q^{2n+1} U \cdot \partial_p^{2n+1} W \qquad (1)$$

which is governing the evolution of the Wigner quasi-probability density $W(p,q,t)$. The traditional Liouville equation, being a milestone of classical statistical mechanics [5], follows from Eq. (1) in the classical limit $\hbar \to 0$. The structure of the Wigner-Moyal equation hints already the stochastic origin of quantum mechanics. In stochastic theory [6], Eq. (1) is a particular example of the well-known Kramers-Moyal equation [4]

$$\partial_t W + p \cdot \partial_q W / m = \sum_{n=1}^{\infty} \frac{(-1)^n}{n!} \partial_p^n \cdot (\alpha_n W) \qquad \alpha_n \equiv \lim_{\Delta t \to 0} \frac{<(\Delta p)^n>}{\Delta t} \qquad (2)$$

The functions $\alpha_n = -(i\hbar/2)_{\text{Re}}^{n-1} \partial_q^n U$ represent jump moments in the momentum subspace, which are also known as the Kramers-Moyal coefficients [6]. It is evident from their definition in Eq. (2) that the higher jump moments describe stochastic processes non-differentiable in the common sense. Since quantum mechanics is time reversible, there are no diffusion terms in Eq. (2), because $\alpha_n \equiv 0$ for any even $n$. While $\alpha_1$ reflects the classical Newton equation, the existence of higher jump moments $\alpha_{n>2} \neq 0$ indicates non-differentiable trajectories of quantum particles in the momentum subspace [7]. The Pawula theorem [6] states that either $\alpha_n \equiv 0 \; \forall n \geq 3$ or all jump moments are meaningful. It follows from positivity of the probability density, which is, however, not the case of the Wigner quasi-probability density $W$ in quantum mechanics. Nevertheless, the Pawula theorem imposes some restrictions on the external potential: $U(q)$ could be constant, linear, harmonic or a general function with infinite number of $q$-derivatives. For the particular potentials above the corresponding Eq. (2) is purely classical, since $\alpha_n \equiv 0 \; \forall n \geq 2$. The negativity problem of the Wigner distribution is probably due to use of improper potentials in quantum mechanics or mathematically needed truncations of the infinite sum of Eq. (1).

In classical mechanics, the Hamilton function defines a system in mechanical sense and it governs the whole evolution of the particles momenta and coordinates. Since $H$ is a sum of the particles kinetic and potential energies, one can generalize further Eq. (2) in a dual Kramers-Moyal form, symmetric on the particles momenta and coordinates,

$$\partial_t W = \sum_{n=1}^{\infty} \frac{(-1)^n}{n!} [\partial_p^n \cdot (\alpha_n W) + \partial_q^n \cdot (\beta_n W)] \qquad \beta_n \equiv \lim_{\Delta t \to 0} \frac{<(\Delta q)^n>}{\Delta t} \qquad (3)$$

The jump moments here are given by $\alpha_n = -(i\hbar/2)_{\text{Re}}^{n-1} \partial_q^n H$ and $\beta_n = (i\hbar/2)_{\text{Re}}^{n-1} \partial_p^n H$ [4]. It is evident from the last expression in Eq. (3) that the higher jump moments describe non-differentiable stochastic processes. For instance, the jump moment $\beta_2 = \hbar/m$ describes the Wiener process used by Nelson [1]. The first term $\partial_q H \cdot \partial_p W - \partial_p H \cdot \partial_q W$ in the sum above is the classical Poisson bracket. Usually, for $H = p^2/2m + U$ the particle trajectory is differentiable in the coordinate subspace, since $\beta_1 = p/m$ and the higher jump moments $\beta_{n>1} = 0$ equal to zero. This is not the case, however, for a relativistic quantum particle, whose Hamilton function reads $H = Mc^2 + U$.

Since the Einstein relativistic mass $M \equiv (m^2 + p^2/c^2)^{1/2}$ depends on the particle momentum, the corresponding jump moments $\beta_n = (i\hbar/2)^{n-1}_{\text{Re}} c^2 \partial_p^n M$ are not zero for any odd $n$. This means that the trajectory of a quantum relativistic particle is not differentiable in the coordinate subspace as well. To examine the natural extension (3) of the Wigner-Moyal equation to relativistic quantum particles, let us consider the simplest case of a free particle. If the latter is also relatively slow, its Hamilton function can be expanded in series to obtain

$$H = Mc^2 = mc^2 + p^2/2m - p^4/8m^3c^2 + \cdots \qquad (4)$$

The corresponding jump moments read $\alpha_n = 0$, $\beta_1 = (1 - p^2/2m^2c^2)p/m$, $\beta_3 = 3(\hbar/2mc)^2 p/m$ and $\beta_{n>3} = 0$. Introducing them in Eq. (3) results in the following equation

$$\partial_t W + p \cdot \partial_q (W - \hat{H}W/mc^2)/m = 0 \qquad (5)$$

where $\hat{H} \equiv p^2/2m - \hbar^2 \partial_q^2/8m$ is the non-relativistic Hamiltonian operator of the free particle in the Wigner phase space representation. Its eigenfunction $\hat{H}W_{st} = EW_{st}$ is the stationary non-relativistic Wigner function, while its eigenvalue $E$ is the non-relativistic particle energy. Thus, the relativistic correction in Eq. (5) seems to be the correct one.

Introducing the partial Fourier transformation in the coordinate subspace, where the Fourier image of $W(p,q,t)$ along the particle coordinate is $W_k(p,t)$, Eq. (5) changes to

$$\partial_t W_k = ik \cdot p W_k (1 - p^2/2m^2c^2 - \hbar^2 k^2/8m^2c^2)/m = ik \cdot p W_k / M_k \qquad (6)$$

It is evident that $M_k = m/(1 - p^2/2m^2c^2 - \hbar^2 k^2/8m^2c^2)$ is the effective mass of the relativistic quantum particle [8]. Since Eq. (6) is derived for a relatively slow particle, the effective mass can be further elaborated to $M_k = m(1 + p^2/2m^2c^2 + \hbar^2 k^2/8m^2c^2)$, which is the power series of

$$M_k = (m^2 + p^2/c^2 + \hbar^2 k^2/4c^2)^{1/2} \qquad (7)$$

It follows from the partial Fourier transformation of the stationary Wigner-Schrödinger equation $\hat{H}W_{st} = EW_{st}$ that the non-relativistic energy of a quantum particle is $E = p^2/2m + \hbar^2 k^2/8m$. Hence, it is straightforward to recognize that Eq. (7) represents the relativistic mass of a quantum particle in the Wigner phase space representation. The Einstein non-quantum relativistic mass $M_{k=0}$ follows in the classical limit $\hbar \to 0$. In the case of a massless particle ($m \equiv 0$) at its lower energy level as a particle ($p = 0$), for instance, the corresponding quantum wave energy is the zero-level vacuum ones, $M_k c^2 = \hbar c k/2 = \hbar\omega/2$. Thus, Eq. (7) properly accounts for the quantum wave-particle dualism.

The Kramers-Moyal equation (3) defines also $\Phi \equiv \sum(\alpha_n \cdot \xi^n + \beta_n \cdot \zeta^n)/n!$ and this kinetic potential contains all information about the jump moments [9]. It is easy to check that the kinetic potential $\Phi = (2/\hbar)H(p + i\hbar\zeta/2, q - i\hbar\xi/2)_{Im}$ [4] generates the jump moments for quantum mechanics $\alpha_n = -(i\hbar/2)_{Re}^{n-1}\partial_q^n H$ and $\beta_n = (i\hbar/2)_{Re}^{n-1}\partial_p^n H$. As is expected, the Hamilton function $H$ determines the kinetic potential of a mechanical system. Thus, Eq. (3) can be also written as $\partial_t W = (2/\hbar)H(p - i\hbar\partial_q/2, q + i\hbar\partial_p/2)_{Im} W$. In the non-relativistic case with $H = p^2/2m + U$, this kinetic potential reduces to

$$\Phi = \zeta \cdot p/m + (2/\hbar)U(q - i\hbar\xi/2)_{Im} \tag{8}$$

In the classical limit Eq. (8) tends to $\Phi = \zeta \cdot p/m - \xi \cdot \partial_q U$. For a free relativistic quantum particle the kinetic potential acquires the form $\Phi = (2c^2/\hbar)M(p + i\hbar\zeta/2)_{Im}$, which reduces to $\Phi = c\zeta$ for a massless particle.

It is important to find out what stochastic process is driving the quantum motion. It is well known that the Schrödinger equation for a free particle $\partial_t \psi = i\hbar\partial_q^2 \psi/2m$ is, in fact, the diffusion equation with imaginary diffusion constant $i\hbar/2m$. Thus, the quantum motion is a Brownian movement with imaginary stochastic force [7]. The classical diffusion equation for the evolution of the probability density $\rho(q,t) = \int W(p,q,t)dp$ of a Brownian particle reads

$$\partial_t \rho = D\partial_q^2 \rho \tag{9}$$

where $D$ is a real diffusion coefficient. The solution of Eq. (9) is a normal distribution density $\rho \sim \exp(-q^2/2\sigma^2)$ with dispersion $\sigma^2 = \sigma_0^2 + 2Dt$ linearly increasing in time. If $D$ is replaced by $i\hbar/4m$, $\rho$ becomes a complex function. Hence, to get the real probability density back, one should take the modulus of the result, which yields again a Gaussian distribution

$$\rho \sim \left|\exp[-q^2/2(\sigma_0^2 + i\hbar t/2m)]\right| = \exp(-q^2/2\sigma^2) \tag{10}$$

Surprisingly, Eq. (10) coincides with the rigorous quantum distribution for spreading of a Gaussian wave packet in vacuum, where $\sigma^2 = \sigma_0^2 + (\hbar t/2m\sigma_0)^2$.

It is interesting how to modify the diffusion equation to be able to describe diffusion with an imaginary diffusion coefficient. A naive possibility is to differentiate Eq. (9) on time to obtain

$$\partial_t^2 \rho = D\partial_q^2 \partial_t \rho = D^2 \partial_q^4 \rho = -\hbar^2 \partial_q^4 \rho / 4m^2 \tag{11}$$

This equation captures already some important features of quantum mechanics. After application of Fourier transformation on time and space, Eq. (11) reduces to the standard quantum dispersion relation $\hbar\omega = (\hbar k)^2/2m$. Another more sophisticated way to describe imaginary diffusion is to introduce the local Shannon information $S \equiv -mD\ln\rho$, measured in action units. Using the entropy $S$, the diffusion equation (9) can be transformed to the following system of two differential equations [10]

$$\partial_t \rho = -\partial_q \cdot (\rho \partial_q S/m) \qquad \partial_t S + (\partial_q S)^2/2m + Q = 0 \tag{12}$$

where $Q \equiv 2mD^2 \partial_q^2 \rho^{1/2}/\rho^{1/2}$. Since the diffusion coefficient appears here on quadrate, one can replace $D$ directly by $i\hbar/2m$ to obtain the Bohm quantum potential $Q = -\hbar^2 \partial_q^2 \rho^{1/2}/2m\rho^{1/2}$. Thus, Eq. (12) reduces to the Bohmian mechanics [11], which is mathematically equivalent to the Schrödinger equation with wave function $\psi \equiv \rho^{1/2}\exp(iS/\hbar)$. Using the latter, the standard momentum operator $\bar{\psi}\hat{p}\psi/m = \rho\partial_q S/m - i\hbar\partial_q \rho/2m$ results in the real convective Bohm flow and the imaginary diffusive Fick flux. Therefore, the Bohm hidden variables are, in fact, imaginary

Wiener processes, due to collisions of the observable real particle with virtual particles, being the hidden carriers of the mechanical forces [12]. Hence, the infinite instant velocities belong to the virtual particles, not to the real ones, as in the Nelson description is [1].

It is important to find out what stochastic process is driving the quantum motion. The stochastic equation proposed by Nelson possesses a drift term, depending on the probability density $\rho$ [1]. Similar to the Bohmian mechanics, this is an indication for a mean-field approach. To resolve the problem between the Brownian motion irreversibility and time reversible quantum mechanics, Nelson considered unphysical evolution back and forward in time. However, how it is discussed above, the Schrödinger equation resembles diffusion with imaginary diffusion constant $i\hbar/2m$, which assures also time reversibility. It is proposed in a previous paper [12], quantum mechanics emerges from the stochastic dynamics of virtual force carriers. Thus, the imaginary diffusion coefficient describes Brownian motion of real particles in the sea of virtual ones. Therefore, the stochastic noise should be purely imaginary one. A possibility to combine the latter with the real Newtonian dynamics is given by

$$m\ddot{R} = \frac{U(R - i\hbar\xi/2) - U(R + i\hbar\xi/2)}{i\hbar\xi} = \frac{U(R - i\hbar\xi/2)_{\text{Im}}}{\hbar\xi/2} \tag{13}$$

where $R(t)$ is the real trajectory of the quantum particle and $\xi(t)$ is a real noise. Here the potential $U$ is of general type, because in the particular cases of constant, linear or harmonic potentials Eq. (13) is not stochastic. Note the correspondence of the last term in Eq. (13) with Eq. (8). Expanding the right-hand side of Eq. (13) in series of $\xi$ results in an alternative form

$$m\ddot{R} = -\sum_{n=0}^{\infty} \frac{(i\hbar/2)^{2n}}{(2n+1)!} \partial_R^{2n+1} U \cdot \xi^{2n} \tag{14}$$

As expected, Eq. (14) reduces to the Newton equation $m\ddot{R} = -\partial_R U$ in the classical limit $\hbar \to 0$. Introducing a positively defined phase space probability density via $W \equiv <\delta(p - m\dot{R})\delta(q - R)>$ and differentiating it on time yields

$$\partial_t W + p \cdot \partial_q W / m = -\partial_p \cdot < m\ddot{R}\delta(p - m\dot{R})\delta(q - R) > \tag{15}$$

Substituting now here the acceleration from Eq. (14) results in

$$\partial_t W + p \cdot \partial_q W / m = \sum_{n=0}^{\infty} \frac{(i\hbar/2)^{2n}}{(2n+1)!} \partial_q^{2n+1} U \cdot \partial_p <\xi^{2n}\delta(p-m\dot{R})\delta(q-R)> \qquad (16)$$

This equation coincides with the Wigner-Moyal equation (1), if the last statistical moment is given by $<\xi^{2k}\delta(p-m\dot{R})\delta(q-R)> \equiv \partial_p^{2k} W$. From this relation one can easily derive $<p^2\xi^2> = 2$, which reveals that $\hbar\xi/2$ is the random fluctuation of the coordinate of the incident force carrier [11]. Since the latter is a virtual particle, its coordinate is multiplied by the imaginary unit in Eq. (13).

The paper is dedicated to the 100$^{th}$ anniversary of David Bohm (1917-1992).

[1]  E. Nelson, *Phys. Rev.* **150** (1966) 1079

[2]  E. P. Wigner, *Phys. Rev.* **40** (1932) 749

[3]  H. J. Groenewold, *Physica* **12** (1946) 405

[4]  J. E. Moyal, *Math. Proc. Cambridge Phil. Soc.* **45** (1949) 99

[5]  J. W. Gibbs, *Proc. Am. Assoc. Adv. Sci.* **33** (1884) 57

[6]  H. Risken and T. Frank, *The Fokker-Planck Equation*, Springer, Berlin, 1996

[7]  D. Prodanov, *J. Phys. Conf. Ser.* **701** (2016) 012031

[8]  R. Tsekov, *Ann. Univ. Sofia, Fac. Phys.* **105** (2012) 22

[9]  R. L. Stratonovich, *Nonlinear Non-equilibrium Thermodynamics*, Nauka, Moscow, 1985

[10] E. Heifetz, R. Tsekov, E. Cohen and Z. Nussinov, *Found. Phys.* **46** (2016) 815

[11] D. Bohm, *Phys. Rev.* **85** (1952) 166

[12] R. Tsekov, *J. Phys. Conf. Ser.* **701** (2016) 012034